# FRAGMENTATION BRANCHING RATIOS OF HIGHLY EXCITED HYDROCARBON MOLECULES $C_nH$ AND THEIR CATIONS $C_nH^+$ ($n \leq 4$)

**T.Tuna [1], M.Chabot [1], T.Pino [2], P.Désesquelles [3], A.LePadellec [4], G.Martinet [1], M.Barat[5], B.Lucas [5], F.Mezdari [6], L.Montagnon [4], N.T. Van-Oanh [7], L.Lavergne [1], A.Lachaize [1], Y. Carpentier [2] and K.Béroff [5]**

1: Institut de Physique Nucléaire, CNRS-IN2P3 and Université Paris-Sud, 91405 Orsay Cedex, France

2: Laboratoire de Photo-Physique Moléculaire, CNRS and Université Paris-Sud, bâtiment 210, F-91405 Orsay Cedex, France

3: Centre de Spectrométrie Nucléaire et Spectrométrie de Masse, Université Paris-Sud and CNRS-IN2P3, bâtiment 104, 91405 Orsay Cedex, France

4: Institut de Recherche sur les Systèmes Atomiques et Moléculaires Complexes, Université Paul Sabatier and CNRS, bâtiment 3R1B4, 31062 Toulouse Cedex 9, France

5: Laboratoire des Collisions Atomiques et Moléculaires, CNRS and Université Paris-Sud, bâtiment 351, 91405 Orsay Cedex, France

6: Institut des Nanosciences de Paris, Université Paris 6 and CNRS, Campus Boucicaut, 140 rue de Lourmel, 75015 Paris Cedex, France

7: Laboratoire de Chimie Physique, Université Paris-Sud and CNRS, bâtiment 349, 91405 Orsay Cedex, France

## Abstract

We have measured fragmentation branching ratios of neutral $C_nH$ and $C_nH^+$ cations produced in high velocity (4.5 a.u) collisions between incident $C_nH^+$ cations and helium atoms. Electron capture gives rise to excited neutral species $C_nH$ and electronic excitation to excited cations $C_nH^+$. Thanks to a dedicated set-up, based on coincident detection of all fragments, the dissociation of the neutral and cationic parents were recorded separately and in a complete way. For the fragmentation of $C_nH$, the H-loss channel is found to be dominant, as already observed by other authors. By contrast, the H-loss and C-loss channels equally dominate the two-fragment break up of $C_nH^+$ species. For these cations, we provide the first fragmentation data ($n > 2$). Results are also discussed in the context of astrochemistry.

## I-INTRODUCTION

Hydrocarbon radicals $C_nH$ have been observed in a number of environments in the interstellar medium : diffuse clouds[1], cold dense molecular clouds and star forming regions[2], photodissociation regions[3,4] and circumstellar envelopes[5]. Mostly neutral species have been detected, with the exception of recently discovered $C_nH^-$ anions[6]. However, cations are included in astrochemical models[7]. Fragmentation of these species under UV photons[3], cosmic rays[8] and dissociative recombination [9] will play a major role in the ongoing chemistry. For all of these processes, parent molecules are formed in electronically excited states before they decay into daughter fragments.

To our knowledge, all measurements concerning the fragmentation of electronically excited $C_nH^{(+)}$ species have been performed with impinging fast or slow electrons. In measurements based on crossed $C_nH^+$ (n = 1,2) ions with fast electron beams, some charged fragment production cross sections have been reported over a broad energy domain [10-12]. These cross sections result generally, with some exceptions [12], from the contribution of both the dissociative excitation and ionization of the parent. Mass spectra of charged fragments following impact of 70 eV electrons on $C_4H$ [13] and $C_5H$ [14] have also been reported. They result from the sum of single and double ionization. Recent measurements of dissociative recombination branching ratios have been conducted at storage rings on $C_2H^+$ [15], $C_3H^+$ [16] and $C_4H^+$ [17]. Contrary to previous measurements, the process of dissociative recombination is well isolated in these experiments.

From the theoretical side, studies concerning the structure of neutral [18,19] and cationic[20] hydrocarbon molecules have been conducted. In recent works, electronic excited states above the dissociation threshold have been computed for $C_nH$ (n = 4-6) [21,22] and $C_5H^+$ [23]. In the case of $C_4H$, energies of all dissociation reactions were calculated using multireference MRSCF calculations with ZPE (zero point energy) corrections [21]. However, this type of information is relatively rare in the literature. Indeed, dissociation energies need consistent calculations and are very sensitive to the level of the quantum calculations. With that respect, information brought by branching ratio measurements, often connected to the values of dissociation energies, may be very useful.

In this work, the excitation and fragmentation of small $C_nH^+$ molecules (n = 1-4) colliding with helium atoms at high velocity (v = 4.5 atomic units) has been investigated experimentally. In this fast collision (~$10^{-16}$ s), the excitation is exclusively of electronic origin. Cross sections of various electronic processes (excitation, ionisation, electron capture) have been measured and the subsequent fragmentation totally recorded. The experimental technique, similar to the one used for the study of small carbon clusters [24], has been adapted to the detection of light (H) and hydrogenated ($C_pH$) fragments. With appropriate modifications, we could extract results at the same level of precision than previously done, *i.e* measurements of all individual branching ratios associated with each dissociative process. In this paper, only the fragmentation of neutral and singly charged molecules will be presented. Apart from its own interest, this experimental study is expected to bring valuable information to be used in an astrochemical context, as discussed later. The plan of the paper is the following. In section II, we describe the experimental method. Results of branching ratios are presented and discussed in section III. In section IV they are discussed in the context of astrochemistry.

## II-EXPERIMENTS

The experiments have been performed at the Tandem accelerator in Orsay. The set-up is composed of three parts: the source of $C_nH^-$ anions, the $C_nH^+$ production system and the AGAT (AGrégat-ATome) set-up where the collision is performed and analysed.

### II-1 The anion source

The anions were produced by a sputtering source: a 20 keV $Cs^+$ beam impinged onto a rod made of 90% of coronene ($C_{24}H_{12}$) and 10% of Ag. Measured intensities of the anions at the exit of the source showed that mostly carbon clusters are formed. A small proportion of $C_nH^-$ molecules (~10%) and a smaller proportion of $C_nH_2^-$ species (~1%) were also produced.

The fact that we obtained a low hydrogenation of the molecules is due to the temperature of the source. The latter could be estimated, as in the case of sputtering from a graphite rod [25], to 3500 K. Beams of $C_nH^-$ are contaminated by $^{13}C^{12}C_{n-1}$ clusters, of equal mass. Since intensities of $C_n^-$ peaks were much larger than those of $C_nH^-$, the contamination of these latter by $^{13}C^{12}C_{n-1}$ beams was important. Taking into account the 1% natural abundance of $^{13}C$ with respect to $^{12}C$, we found that this contamination varied between 15% and 50% depending on n. After subtraction of this contamination, we get $C_nH^-$ intensities around 5 nA, 6 nA, 200 pA and 900 pA for n = 1,2,3,4 respectively. The fact that even n species were found more abundant is in agreement with measurements from other authors [19]. It is explained by a higher electron affinity of the even n species.

### II-2 Stripping and selection of $C_nH^+$ species

Mass selected $C_nH^-$ molecules were injected at 0.2MeV in the Tandem accelerator. They were accelerated up to the high voltage terminal where they passed through a low pressure $N_2$ gas cell. The $C_nH^+$ cations, formed by the loss of two electrons in the gas cell, were accelerated again in the second part of the accelerator, magnetically selected and sent towards the experimental set-up AGAT. The high voltage was adjusted for each value of n in order to have beams with the same velocity of 4.5 atomic units. At this velocity, the flight time between the terminal and the AGAT set-up is 2μs.

The $C_nH^+$ molecules possessed some internal energy, coming from the temperature of the source and the stripping process. The thermal internal energy of $C_nH^-$ was calculated within the canonical ensemble for T=3500 K [26]. In the stripping process, the energy deposit comes from the fact that ejection of electrons from various outer valence shells of $C_nH^-$ can occur. As to ejection of inner valence electrons, it leads to fragmentation of the molecule. In the case of carbon clusters, we estimated the energy deposit due to outer valence electron ionization to 1.5±0.5 eV whatever n, using photoemission spectra of $C_n$ [25]. For $C_nH$ species, due to the lack of such quantities, we assumed the same energy deposit. Altogether we found mean internal energies of $C_nH^+$ of the order of 2 eV for n = 1, 3 eV for n = 2-3 and 4 eV for n = 4 with standard deviations of 1eV for n = 1 and 2 eV for n = 2-4.

With such internal energies $E_i$, various isomers of $C_nH^+$ are populated. In a first approximation, we used the Boltzmann factor, $\exp(^{-\Delta E/Ei})$, for estimating the population of isomers placed at ΔE above the ground state of $C_nH^+$. For $C_2H^+$, we used MCSCF calculations of the five lowest excited states, all below 2 eV [27] , and found that 48% of the ions were in a linear geometry and 52% in a quasi-linear (bent) geometry. For $C_3H^+$, on the basis of CASSCF and CCSD(T) calculations performed on the two lowest isomers [28], we found that 57% of the ions should be in the linear ground state and 43% in the cyclic ($C_3$-ring) isomer situated 0.8 eV above the ground state. For $C_4H^+$, only information on the two lowest energy isomers is available [17]. According to these authors, the energy of the linear form is greater than that of the bi-cyclic isomer by 8 meV. An equal probability of the linear and bi-cyclic isomer is then deduced.

### II-3 The AGAT set-up

The AGAT set-up has already been described [25]. This set-up is made of three parts: the collision chamber, the fragment's electrostatic deflector and the detection chamber. In the collision chamber, an effusive helium jet with a known profile is operating under single collision conditions and absolute cross sections are derived as explained previously [29]. The electrostatic deflector is made of two parallel plates, 15 cm long and 3 cm apart. The electric field was set to 25kV/cm in order to separate fragments of various charge over mass ratios (q/m).

### II-4 Detection of fragments and resolution of the fragmentation channels

In the detection chamber, seven silicon solid-state detectors operating in coincidence intercepted all $C_nH_p{}^{(q+)}$ fragments (n = 0-4, p = 0-1,q = 0-3), see Table I. Three detectors (numbered 1, 2, 3) were "home made" silicon surface barrier detectors mounted at the Institut de Physique Nucléaire in Orsay. They were made of N-type epitaxial silicon of large resistivity (7 kΩ*cm) grown on $N^+$ type silicon substrate with a gold entrance layer (40nm) for the p+ N junction. Their areas were 36mm$^2$ for detector 1 and 72mm$^2$ for detectors 2 and 3. We used fast current preamplifiers designed also at the IPNO. Three detectors (numbered 4,5,6) were commercial ion-implanted silicon detectors of surface area 600mm$^2$ equipped with commercial charge preamplifiers. The detector for $H^+$ fragments was a commercial ion-implanted detector of large dimensions (50 mm x 50 mm) made of 4 independent parallel bands equipped with four commercial charge preamplifiers.

In the standard utilization of silicon solid state detectors, the charge delivered by the detector provides the kinetic energy of the fragment *i.e* its mass for fragments of constant velocity[30]. With this property and the (q/m) analysis, the charge of each fragment was recorded. Specific electronic processes could then be recorded separately. Indeed, charge transfer gives rise to events with only neutral fragments, electronic excitation to events with only one singly charged fragment, single ionization to events with two singly charged fragments or one doubly charged fragment and so on. It has to be noted that non dissociative processes were also recorded, with the exception of the non-dissociative electronic excitation. In the latter case, the excited ion remains in the incident beam and cannot be separated.

Recently we showed that analysing the shape of transient currents allowed resolving, in number and in mass, pile up of many fragments in a detector [31]. This technique, applied successfully to the fragmentation of carbon clusters [29], does not work for H fragments. The reason is that the H fragment, of much lower energy than other fragments, does not contribute significantly to the signal. Therefore, a CH molecular fragment or two atomic C, H fragments give rise to current signals having almost the same shape within the present level of detector noise. Illustration of this is given in figure 1**,** which shows a two-dimensional representation of signals delivered by the neutral detector. The x axis represents the integral of the current signal (proportional to the neutral mass), the y axis its peak value [31]. Each point refers to an event. In this figure, brackets indicate summed signals: {C/H} means two C/H fragments or one CH fragment; C/{C/H} means three C/C/H fragments or two C/CH fragments ….

In order to overcome this problem, we added to the shape analysis the so-called grid method. The grid method was introduced a long time ago in order to resolve the fragmentation of fast $H_3$ molecules [32]. It is still used in dissociative recombination studies carried out at storage rings[33]. The method consists in placing in front of the detector a grid of known transmission and to record mass spectra with and without the grid. The problem is solvable if the number of recorded intensities is equal to or exceeds the number of unknown branching ratios. We applied this technique and placed a grid (hole sizes and wire diameters 20µ) in front of the neutral detector where pile up is involved. The transmission of the grid, t, was measured by recording the attenuation of intensity from a radioactive alpha source ($^{241}$Am) and also by recording the fragmentation pattern of $C_4$ clusters with and without grid. Both methods agreed and we could extract a precise determination of t = 0.269 (± 0.004). As mentioned before, mass spectra and current shapes obtained with and without grid are used for extraction of branching ratios. As an illustration of the method, the dissociation of $C_2H$ is treated in annex.

## III-RESULTS AND DISCUSSION

### III-1-Cross sections of electronic processes

Cross sections of electronic processes measured in the collision of $C_nH^+$ projectiles with helium are reported in figure 2 as a function of n. The ionization process always dominates, followed by dissociative electronic excitation, and electron capture. The relative importance of these processes is as expected in this velocity range [34,35].

The electronic excitation and ionization cross sections exhibit the same evolution with n. This trend is well reproduced by the Independent Atom and Electron (IAE) model [30,36] in the case of ionization. In the IAE calculations, shown by a broken line, classical trajectory Monte Carlo (CTMC) impact parameter probabilities P(b) were used for valence ionization in C and $C^+$ as in previous works [25,37]. Calculations within the SCA (semiclassical approximation) model [38] were performed for P(b) shapes due to inner shell ionization in C and $C^+$ and to ionization of H. Normalizations of P(b) distributions were performed on measured ionization cross sections in $C^+$ [35]and H [39].

Electron capture cross sections evolve in a smoother way with n. The slight increase as a function of n was also observed with incident carbon clusters[24]. It was tentatively interpreted by an increase of the cluster electronic density of states with its size.

### III-2-Relaxation of $C_nH$ species

Tables II, III, IV and V present measured branching ratios for all dissociative and non dissociative relaxation channels of CH, $C_2H$, $C_3H$ and $C_4H$ respectively. Dissociation energies are also reported. In column 3, they were derived by using binding energies of H and C calculated by Pan et al.[19]. In this work, authors used DFT formalism and found ground states to have a quasi-linear geometry. Their dissociation energies were obtained without consideration of symmetry. With the use of symmetries reported in Table VI, we determined - and commented through footnotes in Tables II-V- cases where dissociation from the electronic ground state of the parent to electronic ground states of fragments is forbidden according to correlation rules[40]. When dissociation was forbidden, we considered first electronic excited states of the parent and/or fragments.

#### *a-Number of emitted fragments $N_f$ and internal energy distribution*

Figure 3 shows, for all species, probabilities of dissociation into a given number of fragments $P(N_f)$. These have been obtained by summing branching ratios of Table II, III, IV and V for each $N_f$. We observe, apart from CH, similar probability distributions peaked at two-fragment break up ($N_f$=2).

The number of emitted fragments reflects the internal energy of parent molecules. Indeed, as seen from dissociation energies reported in Tables III to V, a two, three, four, fragments break-up requires in average 6 eV, 12 eV, 19 eV respectively. These formation energies are rather independent of the molecule size. In fact, as we have shown recently on carbon clusters, dissipation in fragments kinetic motion has to be taken into account to correctly connect internal energy and number of emitted fragments. We have:

$$P(Nf) = \int_{E_1}^{E_2} f(E)dE \,/ \int_{0}^{\infty} f(E)\, dE \qquad (1)$$

where f(E) is the internal energy distribution and boundary $E_1$ and $E_2$ express as:

$E_1(N_f)=E_{diss_low}(N_f) +(N_f-2) *E_{TVR}$ for $N_f \geq 2$ (2)

$E_2(N_f)=E_1(N_f+1)$

$E_1(1)=0$

In (2), $E_{diss_low}(N_f)$ is the lowest dissociation energy for a given number of fragments (from Tables II to V), and $E_{TVR}$ is the average energy dissipated in fragment's kinetic motion (translation, vibration, rotation) . We assumed 1eV for $E_{TVR}$ as in the case of $C_n$ [41,42].

In figure 4 is shown, for $C_4H$, the internal energy distribution extracted from data using a step function or the analytical form $E^{a1}exp(-a_2(E-a_3)^{a4}$ for f(E). This last form, depending on four parameters $a_1$-$a_4$, is known to reproduce the energy distribution due to electron capture in any velocity range [29]. We note that the internal energy distribution is very broad, extending from low energies under the dissociation limit up to large energies, well above the ionization potential of $C_4H$. We obtained a similar result in the case of carbon clusters and explained this broad distribution by the electron capture process itself which, in high velocity ion-atom collisions, tends to populate an extended range of excited states [24,34]. Mean internal energies and standard deviations of f(E), reported in parentheses, were found respectively equal to 10 eV (4 eV), 9 eV (6 eV) and 10.5 eV (5 eV) for $C_2H$, $C_3H$ and $C_4H$.

*b-Branching Ratios for fixed $N_f$*

The number of emitted fragments reflects the energy deposited in the system and may be very different from an experiment to another. Branching ratios for a given $N_f$ value are less sensitive to the energy distribution and reflects more intrinsic properties of the parent molecule. In figure 5 are reported branching ratios normalised to 100% for each $N_f$ value for $C_2H$, $C_3H$ and $C_4H$.

For two-fragment break up ($N_f$=2), we see that in all cases the H emission is the most probable dissociation channel. This is not surprising since this channel is always predicted to have the lowest dissociation energy (Tables III-V). The second most populated channel is the C-loss channel. This is in agreement with dissociation energies, except for $C_4H$. For this molecule though, the ordering of the channels is not well established. We also note that the CH-loss is always a small channel.

### III-3 Fragmentation of $C_nH^+$ species

In Tables VII, VIII, IX and X are reported measured branching ratios for all dissociative channels of $CH^+$, $C_2H^+$, $C_3H^+$ and $C_4H^+$ respectively. Dissociation energies are also reported. These were calculated using (3):

$$E_{diss}{}^{+} = E_{diss} - IP_{parent} + IP_{fragment} \qquad (3)$$

where $E_{diss}$ is the dissociation energy of the corresponding neutral channel

$IP_{parent}$ is the adiabatic ionization potential of the neutral parent $C_nH$

$IP_{fragment}$ is the adiabatic ionization potential of the neutral fragment that is singly charged in the exit channel.

$E_{diss}$ were taken from Tables II-V (third column). IP values were taken from the literature and are reported in Table VI. As in the case of neutrals, we indicated in Tables VII to X cases for which ground state to ground state dissociation is not permitted by correlation rules[40]. Symmetries of the cationic states are given in Table VI. One notes that there are many more forbidden transitions with cations than with neutrals. This is often due to non conservation of the spin.

*a-Number of emitted fragments $N_f$ and internal energy distribution*

Figure 6 presents, for each $C_nH^+$ parent, probabilities of dissociation into a given number of fragments $P(N_f)$. As only the dissociative part of excitation is measured in the experiment, measurements begin at $N_f = 2$. We used the method described in III-2 in order to extract internal energies of $C_nH^+$ cations. We found 18 eV (9 eV), 18 eV (7 eV) and 18 eV (7 eV) for mean internal energies (and standard deviations of f(E)) for $C_2H^+$, $C_3H^+$ and $C_4H^+$ respectively. These values are very close to those obtained with $C_n^+$ clusters [43].

*b-Branching Ratios for fixed $N_f$*

In figure 7 are reported dissociation branching ratios, normalised to 100% for each $N_f$ value, for $CH^+$, $C_2H^+$, $C_3H^+$ and $C_4H^+$.

For two-fragment break up, we remark that the H-loss and C-loss channels are almost equally probable. This is not in accordance with dissociation energies, which are always predicted lower for the H-loss channel. For instance, the $C_2^+$/H and $CH^+$/C dissociation energies differ by as much as 2 eV although they have equal branching ratios. The relatively low branching ratio observed for $C_2^+$/H could come from the forbidden ground-state to ground state dissociation (see footnote in Table VIII). It could also be due to dynamical effects - i.e non statistical- since collisional excitation is more probable on the C-C bond than on the C-H bond according to the number of present electrons.

Another difficulty for interpreting these results is the poor confidence in dissociation energies. The IP used in (3) are not well known. For instance, recent determinations of IP in $C_3$[44] and $C_4$[45] are well below all previous values, in particular measurements of Ramanathan et al. [46].

## IV- ASTROCHEMICAL CONTEXT

### IV-1 Branching ratios in a statistical approach

In astrochemical models, reactions rates – i.e. total cross sections - can be, in most cases, safely estimated. Calculations exist and can rely on numerous experimental measurements. On the contrary, branching ratios of different exit channels are much more difficult to obtain. From the experimental point of view, it is because neutral fragments detection is necessary. From the theoretical point of view, it is because *ab initio* dynamical calculations are very difficult and time consuming for systems composed of more than a few atoms [47].

A “simple” theoretical way to predict branching ratios is to use a statistical approach[48]. In such approach, the dynamics vanishes and the fragmentation is governed only by the amount of deposited energy. Statistical branching ratios can be derived from more or less approximated calculations such as Rice–Ramsperger–Kassel–Marcus (RRKM)[49], Weisskopf[50], Phase Space Theory (PST)[51], Microcanonical Metropolis Monte Carlo (MMMC)[29]. Moreover, if the fragmentation is statistical, measurements of branching ratios with a given process can be extended to other processes.

For the fragmentation of small carbon clusters $C_n$ (n=5-9) in high velocity collisions, the statistical approach has been shown to provide satisfactory results [29,42]. The same conclusion was reached on smaller systems studied with photons[52]. In $C_4$, fragmentation branching ratios measured in high velocity collisions [34] are close to those obtained in dissociative recombination [53]. As to the fragmentation of cationic carbon clusters $C_n^+$ (n=4-10), similar results were found for measured branching ratios in photodissociation and in high velocity collisions [54]. All these results tend to prove that, at least for these carbon species, the statistical approach is valid.

### IV-2 Comparison between fragmentation branching ratios of $C_nH$ species following electron capture in high velocity collisions and dissociative recombination

In Table XI are presented dissociative recombination branching ratios of $C_2H^+$ [15], $C_3H^+$ [16] and $C_4H^+$ [17] measured at storage rings. Only two-fragment channels are reported. Indeed, none of the three-fragment channels were observed in $C_3H$ and $C_4H$ fragmentation, due to detection limitation. For the same reason, some channels were not resolved in the fragmentation of $C_3H$ and $C_4H$ and are presented summed in the Table XI. Also reported in Table XI are two-fragment branching ratios that we obtained in this work.

One notices a general good agreement in the ordering of branching ratios obtained in dissociative recombination and high velocity collision experiments. The channel $C_n$/H is found slightly more populated in high velocity collisions than in dissociative recombination for $C_2H$ and $C_4H$. This can be due to the sensitivity of branching ratios to the shape of the internal energy distribution. Indeed, in dissociative recombination, narrow energy distributions -peaked around the neutral parent's IP- vary from one molecule to another while a broad and constant energy distribution is involved in high velocity collisions. The good agreement between branching ratios obtained in dissociative recombination and high velocity collisions could be explained by a statistical fragmentation of $C_nH$ species.

## CONCLUSION

We have presented results concerning the fragmentation of neutral and cationic $C_nH^{(+)}$ hydrocarbons ($n \leq 4$) electronically excited in high velocity collisions with helium atoms. In the case of neutrals, the H-loss channel was found to be the most probable, followed by the C-loss channel and the CH-loss channel. This ordering follows reasonably well dissociation energies and was already obtained by other authors. In contrast, the H-loss and C-loss channels both dominate two-fragment dissociation in case of $C_nH^+$ cations. This result is unexpected since dissociation energies are always predicted lower for H-loss as compared to C-loss. It could be due to dynamical (non statistical) effects. With that respect, it would be interesting to have results obtained in a different experimental situation, for instance, following excitation by another projectile or photon.

We discussed in the paper the astrochemical interest of these measurements. Branching Ratios for two-fragment break-up of $C_nH$ molecules were compared to results of dissociative recombination obtained by other authors. We observed a general good agreement which could be explained by a statistical fragmentation behaviour in these species, as observed in carbon clusters.

**Acknowledgments:**

The authors thank Hocine Khemliche for critical reading of the manuscript.

| Detector | 1 | 2 | 3 | 4 | 5 | 6 | 7 |
|---|---|---|---|---|---|---|---|
| Fragments | H<br>C<br>CH<br>$C_2$<br>$C_2H$<br>$C_3$<br>$C_3H$<br>$C_4$<br>$C_4H$ | $C_4H^+$<br>$C_4^+$<br>$C_3H^+$<br>$C_3^+$ | $C_2H^+$<br>$C_2^+$<br>$C_4H^{++}$<br>$C_4^{++}$<br>$C_3H^{++}$<br>$C_3^{++}$ | $CH^+$<br>$C^+$ | $CH^{++}$<br>$C^{++}$ | $C^{+++}$ | $H^+$ |

**Table I:**

Fragments collected by detectors in the experiment

| Channel | BR(error) (%) | Ediss (eV)[1] | Ediss (eV)[2] |
|---|---|---|---|
| CH | 64(4) | | |
| C/H | 36(4) | 3.65 | 3.45(0.01) |

**Table II:**

Measured branching ratios (BR) of relaxation channels of CH molecules. In last columns, are reported channel dissociation energies ($E_{diss}$ ).
[1]DFT calculations [19]
[2]experimental value from [55]

| Channel | BR(error) (%) | Ediss (eV)[1] | Ediss (eV)[2] |
|---|---|---|---|
| $C_2H$ | 16.8(8) | | |
| $C_2$/H | 51.6(3) | 5.01 | 5.10 |
| C/CH | 11.9(5) | 7.86 | 7.88 |
| 2C/H | 19.7(3) | 11.51 | |

**Table III:**

Measured branching ratios (BR) of relaxation channels of $C_2H$ molecules. In last columns, are reported channel dissociation energies ($E_{diss}$ ).
[1]using DFT calculations [19]
[2]experimental values from [56] and [57] for $C_2$/H and C/CH respectively

| channel | BR(error)(%) | Ediss (eV)[1] | Ediss (eV)[2] |
|---|---|---|---|
| $C_3H$ | 25(3) | | |
| $C_3/H$[a] | 40(7) | 3.46 | 3.08 (4.00) |
| $C_2H/C$ | 21(4) | 6.21 | 5.60 (5.66) |
| $C_2/CH$ | 1(2) | 7.57 | 6.96 (7.03) |
| $C_2/C/H$ | 6.5(3.5) | 11.22 | |
| 2C/CH | 4(2) | 14.07 | |
| 3C/H | 2.5(2.5) | 17.72 | |

**Table IV:**

Measured branching ratios (BR) of relaxation channels of $C_3H$ molecules. In last columns, are reported channel dissociation energies ($E_{diss}$ ).
[1]using DFT calculations [19]
[2]calculations [58] for the linear l-$C_3H$ isomer (and the cyclic c-$C_3H$ one in parentheses)
a: dissociation towards ground state products possible for the linear isomer l-$C_3H$, when bent[58].

| channel | BR(error)(%) | Ediss (eV)[1] | Ediss (eV)[2] | Ediss (eV)[3] |
|---|---|---|---|---|
| $C_4H$ | 11.8(3) | | | |
| $C_4/H$ | 29.9(3) | 4.87 | 5.71 | 4.52 |
| $C_3H/C$ | 13.3(5.5) | 6.90 | 6.33 | 7.25 |
| $C_2/C_2H$ | 7.9(2.5) | 6.61 | 7.12 | 7.24 |
| $C_3/CH$[a] | ≤3 | 6.71 | 5.88 | |
| $C_3/C/H$[b] | 17.1(6.5) | 10.36 | | |
| $2C_2/H$ | 11.1(3.3) | 11.62 | | |
| $C_2/C/CH$ | 1.5(2) | 14.47 | | |
| $C_2H/2C$ | ≤2 | 13.11 | | |
| $C_2/2C/H$ | 6.9(2.8) | 18.12 | | |
| 3C/CH | ≤0.5 | 20.97 | | |
| 4C/H | 0.5(0.5) | 24.62 | | |

**Table V:**

Measured branching ratios (BR) of relaxation channels of $C_4H$ molecules. In last columns, are reported channel dissociation energies ($E_{diss}$ ).
[1]using DFT calculations [19]
[2]MCSCF multireference calculations for linear $C_4H$ including ZPE (zero point energy) corrections [21]
[3]extracted from [33], by using (second and third values) the IP of $C_4H$ (see Table VI).
a,b: ground state to ground state dissociation is not permitted for these channels but dissociation from the electronically excited state of $C_4H$, $^2\Pi$ (+0.02eV [59]) is possible.

| Species (and g.s symmetry) | IP(error) (eV) | | Cation (and g.s. symmetry) |
|---|---|---|---|
| $H(^2S_g)$ | 13.6(<0.01) | [60] | $H^+$ |
| $C(^3P_g)$ | 11.26(<<0.01) | [60] | $C^+(^2P_u)$ |
| $CH(X^2\Pi, C_{\infty v})$ | 10.64(0.01)* | [55] | $CH^+(X^1\Sigma^+, C_{\infty v})$ |
| $C_2(X^1\Sigma^+_g, D_{\infty h})$ | 11.41(0.3)** | [61] | $C_2^+(X^4\Sigma_g^-, D_{\infty h})$ |
| $C_2H(X^2\Sigma^+, C_{\infty v})$ | 11.61(0.07) | [62] | $C_2H^+(X^3\Pi, C_{\infty v})$ |
| $C_3(X^1\Sigma^+_g, D_{\infty h})$ | 11.61(0.07)** | [44] | $C_3^+(X^2B_2, C_{2v})$ |
| c-$C_3H(X^2B_2, C_{2v})$ | 9.06 | [28] | $C_3H^+(X^1\Sigma^+, C_{\infty v})$ |
| l-$C_3H(^2\Pi, C_{\infty v})$(+0.04eV)[28] | 8.36 | [28] | |
| c-$C_4(X^1A_g, D_{2h})$ | 10.9(0.2)** | [45] | $C_4^+(X^2B_{1u}, D_{2h})$ |
| l-$C_4(^3\Sigma_g^-, D_{\infty h})$(+0.04eV)[63] | 11.0(0.2)** | [45] | |
| $C_4H(X^2\Sigma^+, C_{\infty v})$ | 12 | [17] | $C_4H^+(X^3\Sigma^-, C_{\infty v})$[a] |

**Table VI:**

Ground states and symmetries of neutral (first column) and cationic (third column) species relevant to this study. For $C_3H$ and $C_4$, we reported also linear isomers which are very close in energy from cyclic ground states. In column 2 are reported ionization potentials used in equation (1); * : vertical IP; ** : adiabatic IP.
a: The $C_4H^+$ molecular ion has been poorly studied and, to our knowledge, there is only one reference in which a $^1\Sigma^+$ state for its electronic ground state has been proposed [64]. Ground state investigations have thus been performed using the Gaussian 2003 suite of programs [65]. The calculations were performed at the B3LYP/cc-pTVZ and CCSD(T)/cc-pVTZ levels. Both were in agreement and predicted that the electronic ground state is in fact a $X^3\Sigma^-$ state.

| Channel | BR(error) (%) | Ediss (eV) |
|---|---|---|
| $C^+$/H | 63.5(1) | 4.27[a] |
| C/$H^+$[b] | 36.5(1) | 6.61 |

**Table VII:**

Measured branching ratios (BR) of dissociative channels of $CH^+$ molecules. In column three are reported channel dissociation energies calculated with equation (1) using Table II and Table VI.
a: experimental value: 3.95 eV (±0.03%) [66]
b: dissociation occurs from the electronic excited state $b^3\Sigma^-$ state of $CH^+$ situated +4.5eV above the ground state [11].

| Channel | BR(error) (%) | Ediss (eV) |
|---|---|---|
| $C_2^+/H$ [a] | 16(2) | 4.81 |
| $CH^+/C$ | 14.5(2) | 6.89 |
| $C_2/H^+$ [b] | 2.6(1) | 7.00 |
| $CH/C^+$ | 2.4(2) | 7.51 |
| $C/H/C^+$ | 47.7(2) | 11.16 |
| $2C/H^+$ | 16.8(1) | 13.50 |

**Table VIII:**

Measured branching ratios (BR) of dissociative channels of $C_2H^+$ molecules. In column three are reported channel dissociation energies calculated with equation (1) using Table III and Table VI.

a: ground state to ground state dissociation is not permitted for this channel; dissociation may occur from an electronic excited state of the parent ($^3\Sigma^-$ at +0.8eV [27]) or lead to an excited state of $C_2^+$ ($^2\Pi_u$ at +0.8eV [67]) b: ground state to ground state dissociation is not permitted for this channel but dissociation may lead to an excited state of $C_2$ ($^3\Pi_u$ at +0.1eV [68])

| Channel | BR(error) (%) | Ediss (eV) |
|---|---|---|
| $C_2H^+/C$ | 7.4(2) | 8.76 |
| $C_3^+/H$ | 10.5(1) | 6.01 |
| $C_2H/C^+$ | 4.7(1) | 8.41 |
| $CH/C_2^+$ [a] | 4.6(1) | 9.91 |
| $C_2/CH^+$ | 3.7(0.5) | 9.14 |
| $C_3/H^+$ | 0.8(0.1) | 8.00 |
| $C/CH/C^+$ | 13.6(2.5) | 16.27 |
| $C/H/C_2^+$ | 14.4(1) | 13.56 |
| $C_2/H/C^+$ | 9.2(1) | 13.41 |
| $C_2/C/H^+$ [b] | 4.3(1) | 15.75 |
| $2C/CH^+$ | 9.2(3) | 15.65 |
| $2C/H/C^+$ | 15.1(1.5) | 21.28 |
| $3C/H^+$ [c] | 2.5(1) | 23.62 |

**Table IX:**

Measured branching ratios (BR) of dissociative channels of $C_3H^+$ molecules. In column three are reported channel dissociation energies calculated with equation (1) using Table IV and Table VI.

a: ground state to ground state dissociation is not permitted for this channel but dissociation leading to an excited state of $C_2^+$ ($^2\Pi_u$ at +0.8eV [67]) is permitted b: ground state to ground state dissociation is not permitted for this channel but dissociation leading to an excited state of $C_2$ ($^3\Pi_u$ at +0.1eV [68]) is permitted c: ground state to ground state dissociation is not permitted for this channel but dissociation leading to an excited state of C (for instance $^1D_g$ at +1.2eV [69]) is permitted

| Channel | BR(error) (%) | Ediss (eV) |
|---|---|---|
| $C/C_3H^+$ | 9.2(1.3) | 3.96 |
| $C_4^+/H^a$ | 5.3(0.4) | 3.77 |
| $C_3H/C^+$ | 2.2(0.7) | 6.16 |
| $C_2/C_2H^{+b}$ | 1.6(0.5) | 6.21 |
| $C_3/CH^{+c}$ | 1.9(0.4) | 5.35 |
| $C_2H/C_2^+$ | 0.7(0.5) | 6.01 |
| $C_4/H^{+d}$ | 0.3(0.1) | 6.47 |
| $C_3^+/CH$ | 1.8(0.4) | 6.32 |
| $C/H/C_3^+$ | 15.1(1.2) | 9.97 |
| $C_3/H/C^{+e}$ | 9.5(0.5) | 9.62 |
| $C_2/H/C_2^+$ | 7.0(1.1) | 11.02 |
| $C_2H/C/C^+$ | 2.9(0.8) | 12.37 |
| $C_2/C/CH^+$ | 3.6(0.5) | 13.10 |
| $C_3/C/H^+$ | 1.8(0.1) | 11.96 |
| $2C/C_2H^+$ | 0.7(0.7) | 12.72 |
| $C/CH/C_2^+$ | 0.7(0.7) | 13.87 |
| $C_2/CH/C^+$ | 2.3(0.5) | 13.72 |
| $2C_2/H^{+f}$ | 0.9(0.1) | 13.21 |
| $C_2/C/H/C^+$ | 13.3(1.3) | 17.37 |
| $2C/H/C_2^+$ | 7.1(1.1) | 17.52 |
| $2C/CH/C^+$ | 2.1(0.7) | 20.23 |
| $C_2/2C/H^+$ | 1.7(0.2) | 19.71 |
| $3C/CH^+$ | 1.5(0.3) | 19.61 |
| $3C/H/C^+$ | 6.3(1) | 25.24 |
| $4C/H^+$ | 0.5(0.1) | 27.58 |

**Table X:**

Measured branching ratios (BR) of dissociative channels of $C_4H^+$ molecules. In column three are reported channel dissociation energies calculated with equation (1) using Table V and Table VI.

a: ground state to ground state dissociation is not permitted; dissociation towards excited $C_4^+$ ($^4\Sigma_g^-$, +0.3eV [70]) is possible.

b: ground state to ground state dissociation is not permitted; dissociation towards excited $C_2$ ($^3\Pi_u$, +0.1eV [68]) is possible.

c, e: ground state to ground state dissociation permitted towards cyclic isomer $C_3$ ($^3A'_2$) situated +0.8eV above the linear isomer [58].

d: ground state to ground state dissociation permitted towards l-$C_4$

f: ground state to ground state dissociation is not permitted; dissociation towards excited $C_2$ ($^3\Sigma_g^-$, +0.9eV [68]) is possible.

| Channel | DR | HV collision |
|---|---|---|
| $C_2$/H | 52(4) | 81(5) |
| C/CH | 48(5) | 19(8) |
| $C_3$/H | 66(1.5) | 65(11) |
| $C_2$H/C | } Σ=34(1.7) | 33(6) |
| $C_2$/CH | | ≤2 |
| $C_4$/H | 44(1.2) | 55(6) |
| C/$C_3$H | } Σ=28(2) | 25(10) |
| $C_3$/CH | | ≤5 |
| $C_2$/$C_2$H | 28(2) | 15(5) |

**Table XI:**

Comparison, for two-fragment break-up, between measured branching ratios in dissociative recombination experiments (DR, adapted from [15], [16] and [17]) and in high velocity collisions by electron capture (HV collision, this work); branching ratios and error bars (in parentheses) are given in %.

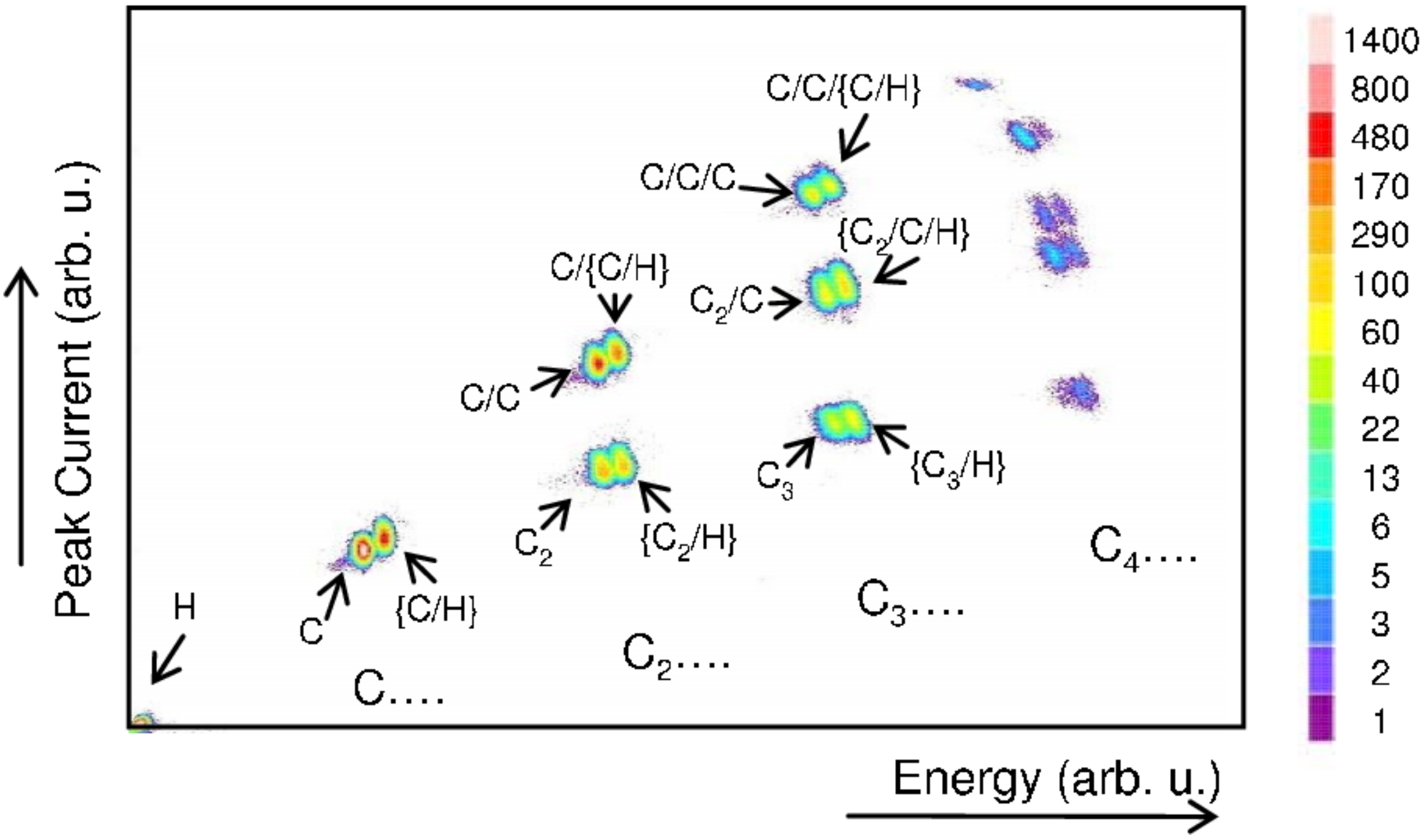

**Figure 1**

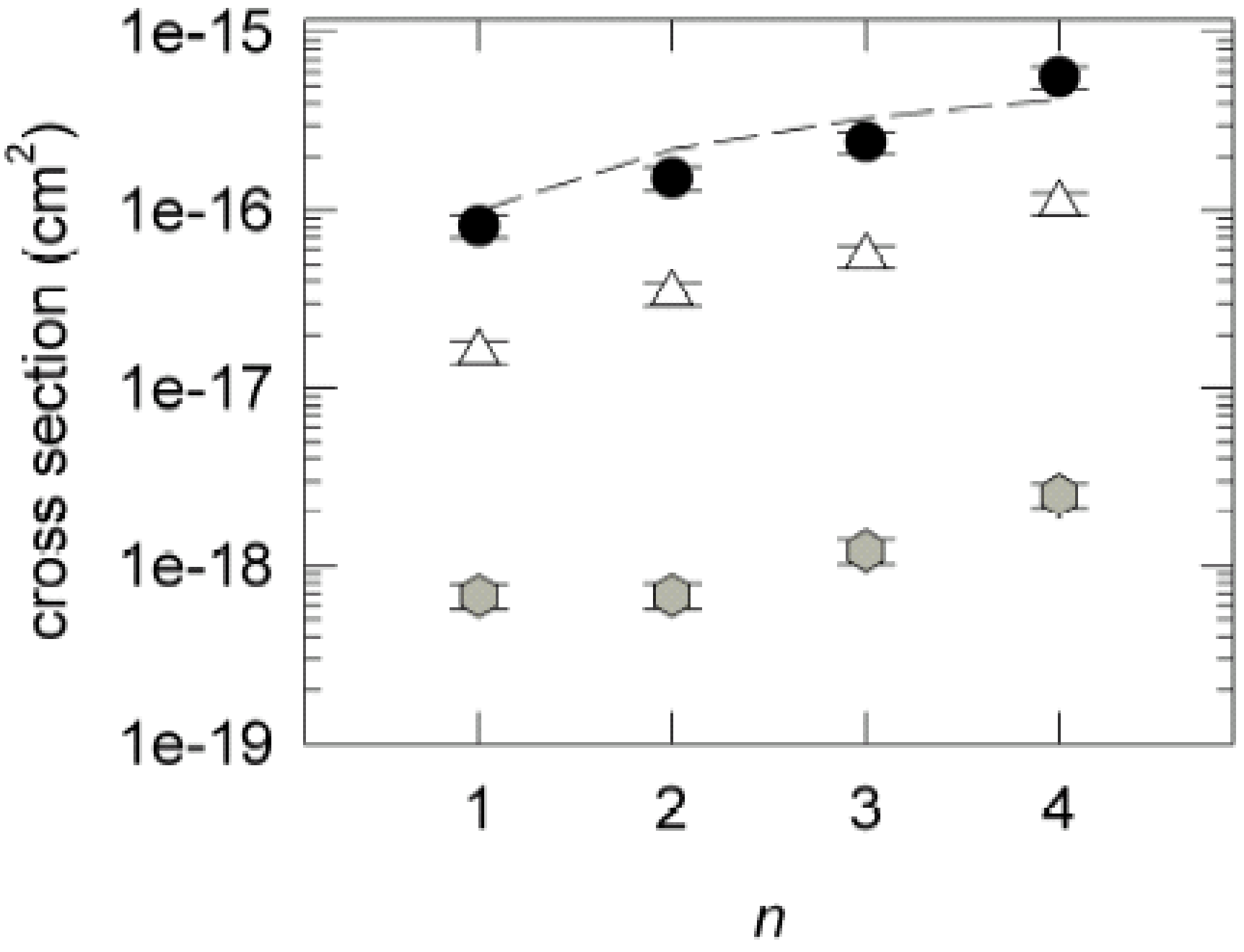

**Figure 2**

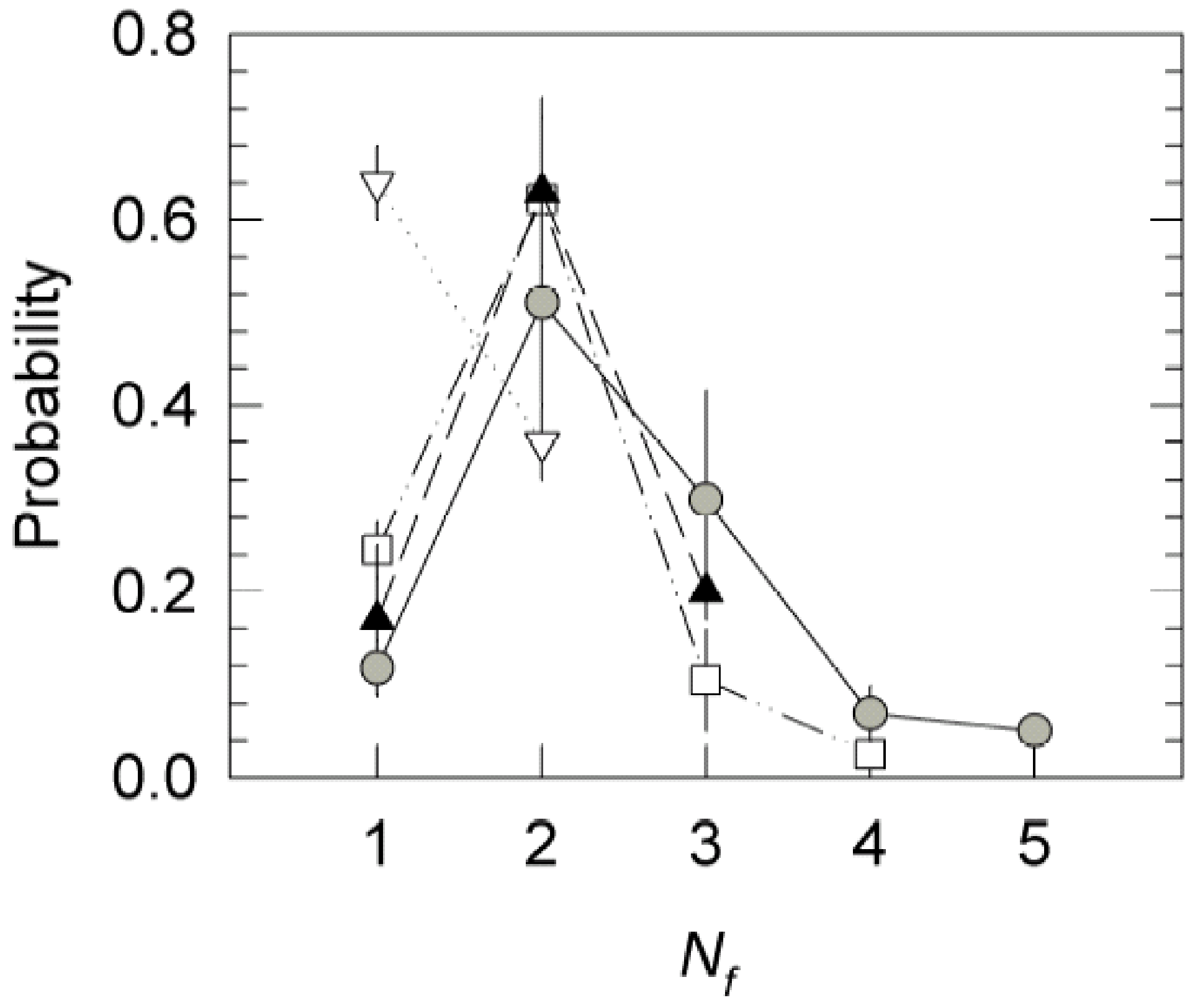

**Figure 3**

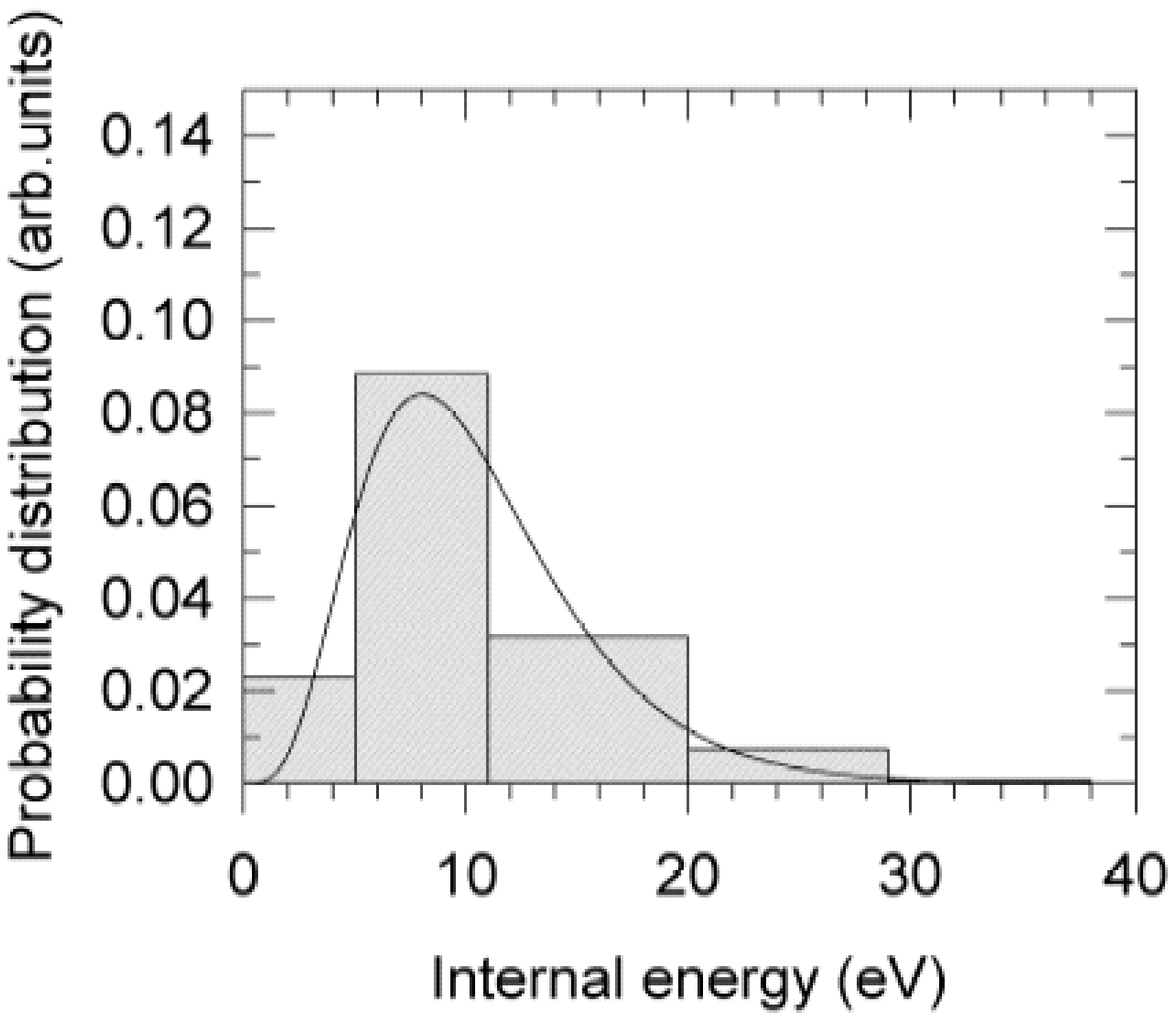

**Figure 4**

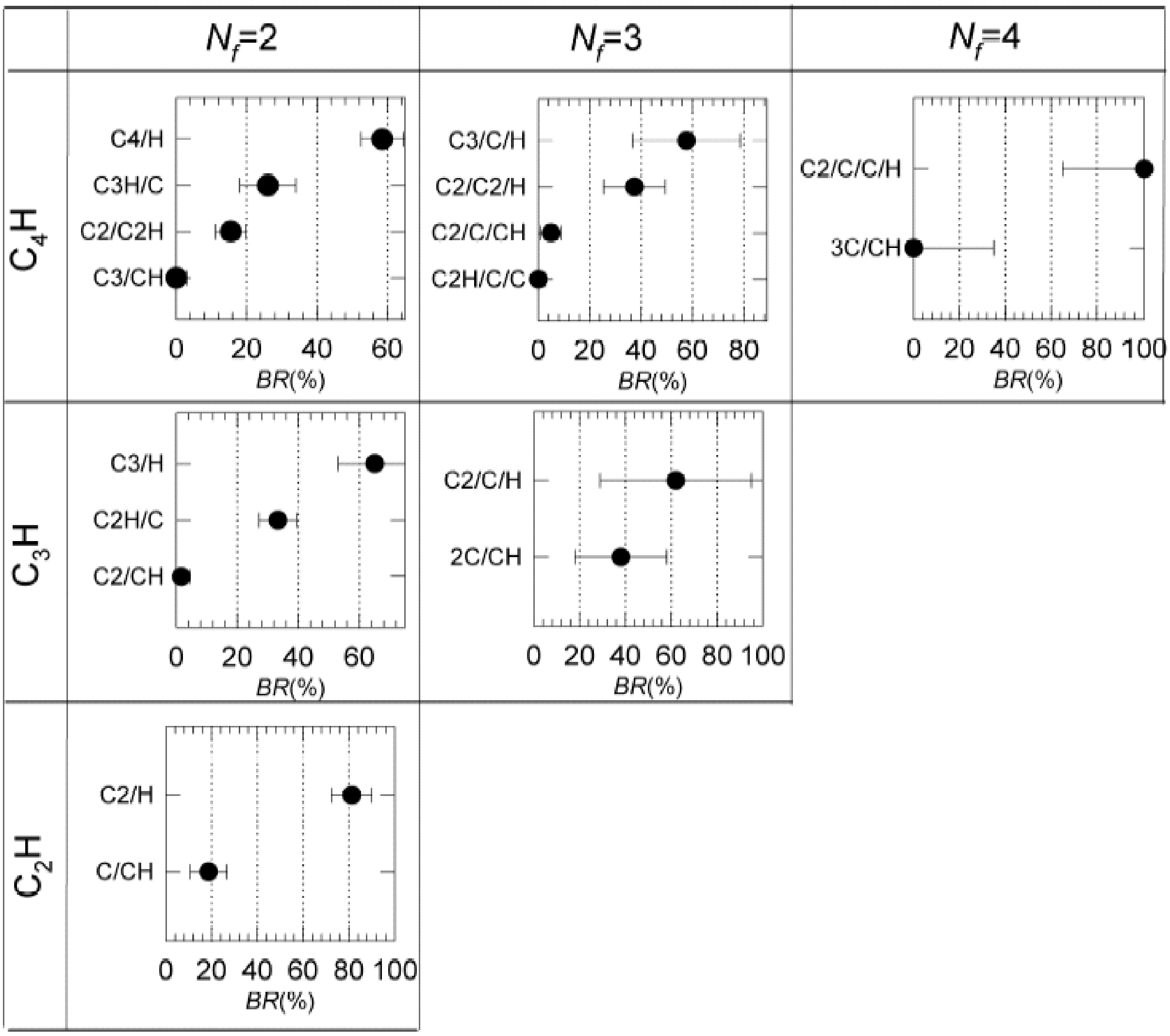

**Figure 5**

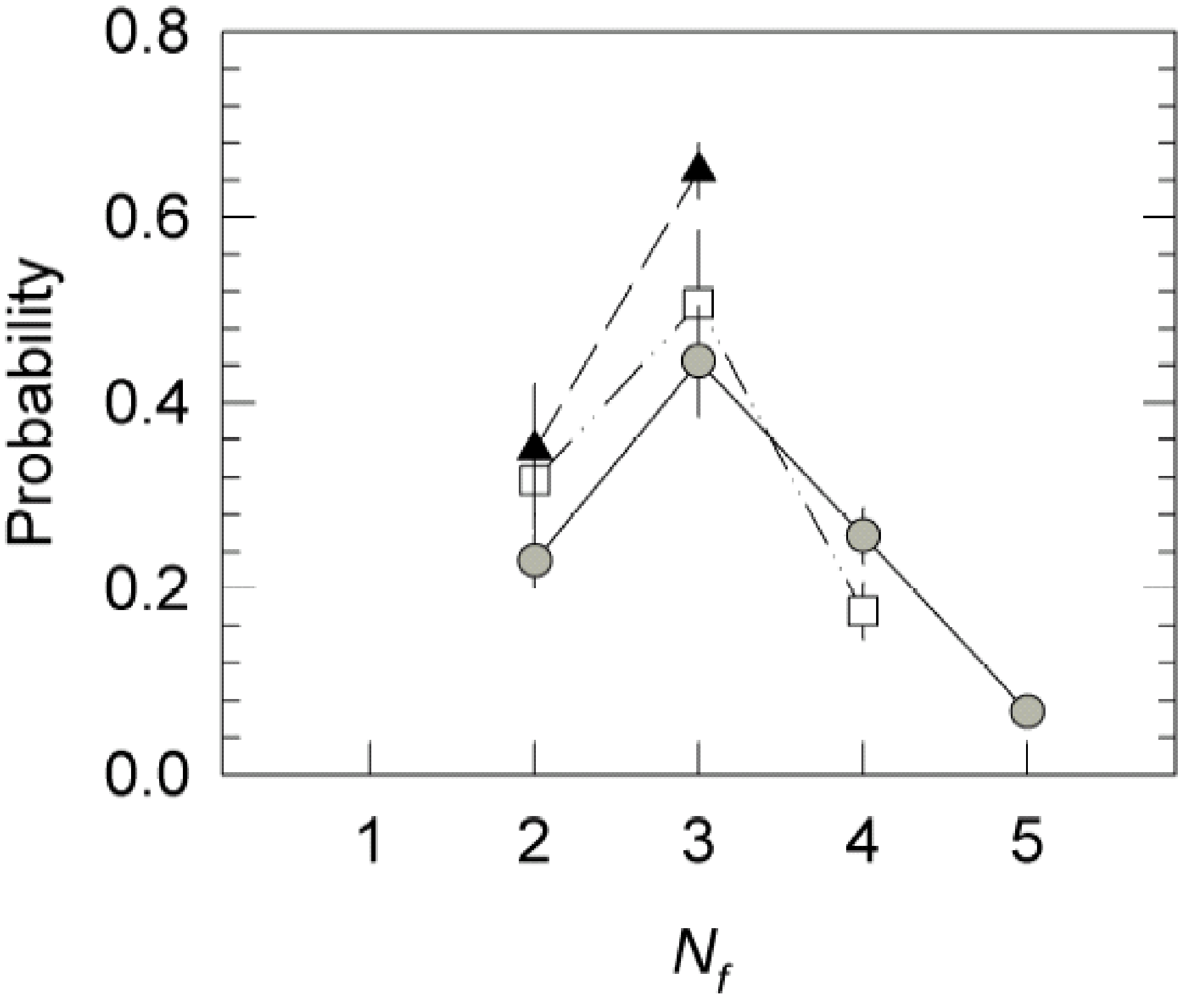

**Figure 6**

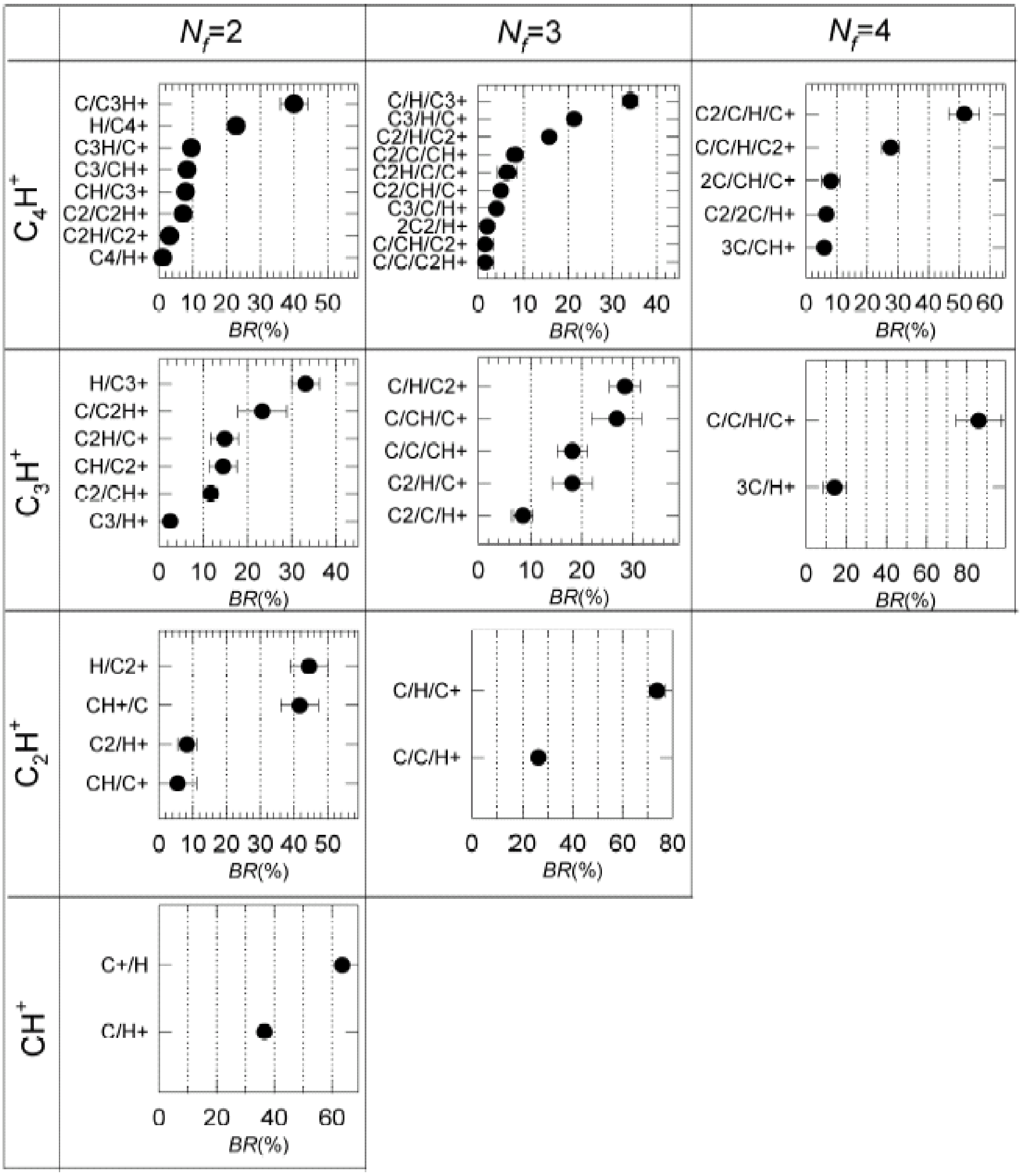

Figure 7

**ANNEX :**

**Extraction of branching ratios by the grid method and current shape analysis**

1- Expression of the transmission matrix; case of $C_2H$ dissociation

A set of linear equations connect the intensities of the measured peaks to the intensities of the various dissociation channels (unknown quantities). We write TX=M where T is the transmission matrix, X is the vector containing the unknown quantities, M the vector containing the measured intensities (with grid and without grid). The probability that a fragment traverses the grid is t, the probability that it is stopped is s = (1-t) and since all fragments are independent, it is easy to write the form of the T matrix.

This is what is done below (equation 1A) for the case of the dissociation of $C_2H$. For the M vector, the index g refers to measurements done with the grid. As to the bracket sign [], it indicates that this measurement is an unresolved sum of two terms (see below). We note that components of the X vector (unknown quantities) contain the four branching ratios of dissociation of $C_2H$ ($C_2H$, $C_2/H$, C/CH and 2C/H) but also channels having an incomplete mass ($C_2$, 2C, CH, C/H, C and H). These incomplete channels come from dissociation of incident $C_2H^+$ on the residual gas and, although small, have to be introduced in the equation especially for estimation of the errors.

$$\begin{pmatrix} t & t^2 & 0 & 0 & 0 & 0 & 0 & 0 & 0 & 0 \\ 0 & 0 & t^2 & t^3 & 0 & 0 & 0 & 0 & 0 & 0 \\ 0 & ts & 0 & 0 & t & 0 & 0 & 0 & 0 & 0 \\ 0 & 0 & 0 & t^2 s & 0 & t^2 & 0 & 0 & 0 & 0 \\ 0 & 0 & ts & 2t^2 s & 0 & 0 & t & t^2 & 0 & 0 \\ 0 & 0 & ts & 2t^2 s & 0 & 2ts & 0 & ts & t & 0 \\ 0 & ts & 0 & ts^2 & 0 & 0 & 0 & ts & 0 & t \\ 1 & 1 & 0 & 0 & 0 & 0 & 0 & 0 & 0 & 0 \\ 0 & 0 & 1 & 1 & 0 & 0 & 0 & 0 & 0 & 0 \\ 0 & 0 & 0 & 0 & 1 & 0 & 0 & 0 & 0 & 0 \\ 0 & 0 & 0 & 0 & 0 & 1 & 0 & 0 & 0 & 0 \\ 0 & 0 & 0 & 0 & 0 & 0 & 1 & 1 & 0 & 0 \\ 0 & 0 & 0 & 0 & 0 & 0 & 0 & 0 & 1 & 0 \\ 0 & 0 & 0 & 0 & 0 & 0 & 0 & 0 & 0 & 1 \end{pmatrix} * \begin{pmatrix} C_2H \\ C_2/H \\ C/CH \\ 2C/H \\ C_2 \\ 2C \\ CH \\ C/H \\ C \\ H \end{pmatrix} = \begin{pmatrix} ([C_2H])_g \\ (C/[CH])_g \\ (C_2)_g \\ (2C)_g \\ ([CH])_g \\ (C)_g \\ (H)_g \\ [C_2H] \\ C/[CH] \\ C_2 \\ 2C \\ [CH] \\ C \\ H \end{pmatrix} \qquad (1A)$$

As mentioned before, the shape analysis method allows to resolve the fragmentation of $C_n$ species. Then, $C_2$ or 2C are separately measured (see components of vector M in equation 1A). On the contrary, the H fragmentation is not resolved. The notation [$C_2$H] in the M vector means that intensities of $C_2$H and $C_2$/H are summed. Similarly [CH] is the sum of CH and C/H intensities and C/[CH] the sum of C/CH and 2C/H intensities.

2-Resolution of equations, branching ratios and error bars

As seen from (1A), 14 measurements (components of the M vector) are performed for 10 unknown quantities (components of the X vector) since t and s are known. Since the system is over-determined, the idea is to test and keep the equations which are the most sensitive to changes in the unknown branching ratios (or equivalently to remove the unsensitive equations which are only bringing uncertainties on the results). The error bars are calculated as follows: measured values are allowed to vary, within experimental error bars, by a Monte Carlo random procedure and unknown intensities of vector X extracted by resolving the equation (1A). The distribution of extracted values for each unknown intensity, obtained by varying the set of measured data, provides directly a mean value and error bar for this intensity. This treatment is performed also for the background (jet not overlapping with the beam), the final error bar taking into account this background subtraction.

3- Subtraction of the $^{13}$C isotope contribution, final branching ratios and error bars

As mentioned in section II, a sizeable contribution of $^{13}C^{12}C_{n-1}^+$ exists in the $C_nH^+$ beams. This means that $^{13}C^{12}C_p^{q+}$ fragments may exist and will appear at the same mass than $C_{p+1}H^{q+}$. For the case of incident $C_2H^+$ beam and restricting to the neutralization (electron capture) process, it means that three new unknown quantities have to be considered in our equations: $^{13}$C (same detector response as CH), $^{12}C^{13}C$ (as $C_2$H) and $^{12}C/^{13}C$ (as C/CH). We could have added these unknown quantities in the equation (1A) and resolve it but we choose to subtract the contribution due to $^{13}$C at the final stage (from the already extracted intensities) since it involves quantities that are measured in the experiment. Indeed we get, for the final intensities:

$$I_{final}\,(C_2H)=I_{extr}(C_2H)-b\ {*}BR(^{13}C^{12}C) \qquad (2A\text{-}a)$$

$I_{final}$ $(C_2/H)=I_{extr}(C_2/H)$ (2A-b)

$I_{final}$ $(C/CH)= I_{extr}(C/CH)- b *BR(^{12}C/^{13}C)$ (2A-c)

$I_{final}$ $(2C/H)=I_{extr}(2C/H)$ (2A-d)

Where $I_{extr}$ are obtained by the procedure described above, b is the percentage of $^{12}C^{13}C^{+}$ intensity in the $C_2H^+$ beam and BR are the branching ratios of de-excitation of excited $^{13}C^{12}C$ created by electron capture in the $^{13}C^{12}C^{+}$-He collision. These quantities are measured in the experiment (recordings with $C_n^+$ beams of identical velocity were systematically performed) and it was assumed that these measurements with $^{12}C_n^+$ beams are applicable to $^{13}C^{12}C_{n-1}^+$ beams (*i.e* there is no isotopic effect on cross section and dissociation patterns).

The b quantity was also measured in the experiment. Indeed, following conclusions of Ben-Itzhak et al [71], we assumed that no $CH^{++}$ ions may survive after 70 ns. Then the peak at M = 13 near the fragment $C^{++}$ (detector 5, see Table I) was assigned to $^{13}C^{++}$ and not to $CH^{++}$ and b was extracted from equation (3A):

$$b=\frac{I_{extr}(^{12}C^{++13}C^{++})}{I_{extr}(M=25,q=4)}*I_{extr}(M=25,q=0) \qquad (3A)$$

Error bars for final branching ratios took into account the error bars for extracted values $I_{extr}$ , the error bar on the b value and error bars on BR values, these last ones quite small since measurements with $C_n^+$ projectiles done without grid have large statistics.

## REFERENCES:

1 R. Lucas and H. S. Liszt, Astronomy and Astrophysics **358** (3), 1069 (2000).

2 P. Thaddeus, C. A. Gottlieb, A. Hjalmarson, L. E. B. Johansson, W. M. Irvine, P. Friberg, and R. A. Linke, Astrophysical Journal **294** (1), L49 (1985); E. F. van Dishoeck and G. A. Blake, Annual Review of Astronomy and Astrophysics **36**, 317 (1998).

3 D. J. Hollenbach and A. Tielens, Annual Review of Astronomy and Astrophysics **35**, 179 (1997).

4 D. Teyssier, D. Fosse, M. Gerin, J. Pety, A. Abergel, and E. Roueff, Astronomy & Astrophysics **417** (1), 135 (2004).

5 M. Guelin, J. Cernicharo, M. J. Travers, M. C. McCarthy, C. A. Gottlieb, P. Thaddeus, M. Ohishi, and S. Saito, Astronomy and Astrophysics **317** (1), L1 (1997); J. R. Pardo and J. Cernicharo, Astrophysical Journal **654** (2), 978 (2007).

6 M. C. McCarthy, C. A. Gottlieb, H. Gupta, and P. Thaddeus, Astrophysical Journal **652** (2), L141 (2006); J. Cernicharo, M. Guelin, M. Agundez, K. Kawaguchi, M. McCarthy, and P. Thaddeus, Astronomy & Astrophysics **467** (2), L37 (2007).

7 E. Herbst, www.physics.ohio-state.edu/~eric/research-files/osu_01_2007 (2007); J. Woodall, M. Agundez, A. J. Markwick-Kemper, and T. J. Millar, Astronomy & Astrophysics **466** (3), 1197 (2007).

8 L. Spitzer and M. G. Tomasko, The Astrophysical journal **152**, 971 (1968); V. Mennella, G. A. Baratta, A. Esposito, G. Ferini, and Y. J. Pendleton, Astrophysical Journal **587** (2), 727 (2003).

9 W. D. Geppert, R. Thomas, J. Semaniak, A. Ehlerding, T. J. Millar, F. Osterdahl, M. A. Ugglas, N. Djuric, A. Paal, and M. Larsson, Astrophysical Journal **609** (1), 459 (2004); W. D. Geppert, R. D. Thomas, A. Ehlerding, F. Hellberg, F. Österdahl, M. Hamberg, J. Semaniak, V. Zhaunerchyk, M. Kaminska, A. Källberg, A. Paal, and M. Larsson, Journal of Physics: Conference Series **4**, 26 (2005).

10 D. B. Popovic, N. Djuric, K. Holmberg, A. Neau, and G. H. Dunn, Physical Review A **64**, 052709 (2001).

11 A. E. Bannister, H. F. Krause, C. R. Vane, N. Djuric, D. B. Popovic, M. Stepanovic, G. H. Dunn, Y. S. Chung, A. C. H. Smith, and B. Wallbank, Physical Review A **68** (4) (2003).

[12] H. Cherkani-Hassani, Thesis, Université de Louvain-la-Neuve (Belgique), unpublished (2004).

[13] X. B. Gu, Y. Guo, and R. I. Kaiser, International Journal of Mass Spectrometry **246** (1-3), 29 (2005).

[14] X. B. Gu, Y. Guo, and R. I. Kaiser, International Journal of Mass Spectrometry **261** (2-3), 100 (2007).

[15] A. Ehlerding, F. Hellberg, R. Thomas, S. Kalhori, A. A. Viggiano, S. T. Arnold, M. Larsson, and M. af Ugglas, Physical Chemistry Chemical Physics **6** (5), 949 (2004).

[16] G. Angelova, O. Novotny, J. B. A. Mitchell, C. Rebrion-Rowe, J. L. Le Garrec, H. Bluhme, A. Svendsen, and L. H. Andersen, International Journal of Mass Spectrometry **235** (1), 7 (2004).

[17] G. Angelova, O. Novotny, J. B. A. Mitchell, C. Rebrion-Rowe, J. L. Le Garrec, H. Bluhme, K. Seiersen, and L. H. Andersen, International Journal of Mass Spectrometry **232** (2), 195 (2004).

[18] T. D. Crawford, J. F. Stanton, J. C. Saeh, and H. F. Schaefer, Journal of the American Chemical Society **121** (9), 1902 (1999).

[19] L. Pan, B. K. Rao, A. K. Gupta, G. P. Das, and P. Ayyub, Journal of Chemical Physics **119** (15), 7705 (2003).

[20] K. Raghavachari, R. A. Whiteside, J. A. Pople, and P. v. R. Schleyer, Journal of the American Chemical Society **103**, 5649 (1981).

[21] S. Graf, J. Geiss, and S. Leutwyler, Journal of Chemical Physics **114** (10), 4542 (2001).

[22] J. Haubrich, M. Muhlhauser, and S. D. Peyerimhoff, Journal of Physical Chemistry A **106** (35), 8201 (2002); Z. X. Cao and S. D. Peyerimhoff, Physical Chemistry Chemical Physics **3** (8), 1403 (2001).

[23] J. Haubrich, M. Muhlhauser, and S. D. Peyerimhoff, Journal of Molecular Structure-Theochem **623**, 335 (2003).

[24] M. Chabot, G. Martinet, F. Mezdari, S. Diaz-Tendero, K. Beroff-Wohrer, P. Desesquelles, S. Della-Negra, H. Hamrita, A. Le Padellec, T. Tuna, L. Montagnon, M. Barat, M. Simon, and I. Ismail, Journal of Physics B-Atomic Molecular and Optical Physics **39** (11), 2593 (2006).

[25] F. Mezdari, K. Wohrer-Beroff, M. Chabot, G. Martinet, S. Della Negra, P. Desesquelles, H. Hamrita, and A. Le Padellec, Physical Review A **72** (3) (2005).

[26] J. U. Andersen, E. Bonderup, and K. Hansen, Journal of Chemical Physics **114** (15), 6518 (2001).
[27] K. Hashimoto, S. Iwata, and Y. Osamura, Chemical Physics Letters **174** (6), 649 (1990).
[28] S. Ikuta, Journal of Chemical Physics **106** (11), 4536 (1997).
[29] G. Martinet, S. Diaz-Tendero, M. Chabot, K. Wohrer, S. Della Negra, F. Mezdari, H. Hamrita, P. Desesquelles, A. Le Padellec, D. Gardes, L. Lavergne, G. Lalu, X. Grave, J. F. Clavelin, P. A. Hervieux, M. Alcami, and F. Martin, Physical Review Letters **93** (6) (2004).
[30] K. Wohrer, M. Chabot, J. P. Rozet, D. Gardes, D. Vernhet, D. Jacquet, S. DellaNegra, A. Brunelle, M. Nectoux, M. Pautrat, Y. LeBeyec, P. Attal, and G. Maynard, Journal of Physics B-Atomic Molecular and Optical Physics **29** (20), L755 (1996).
[31] M. Chabot, S. Della Negra, L. Lavergne, G. Martinet, K. Wohrer-Beroff, R. Sellem, R. Daniel, J. Le Bris, G. Lalu, D. Gardes, J. A. Scarpaci, P. Desesquelle, and V. Lima, Nuclear Instruments & Methods in Physics Research Section B-Beam Interactions with Materials and Atoms **197** (1-2), 155 (2002).
[32] K. H. Berkner, T. J. Morgan, R. V. Pyle, and J. W. Stearns, Physical Review A **8** (6), 2870 (1973); J. L. Forand, J. B. A. Mitchell, and J. W. McGowan, Journal of Physics E-Scientific Instruments **18** (7), 623 (1985).
[33] A. I. Florescu-Mitchell and J. B. A. Mitchell, Physics Reports **430**, 277 (2006).
[34] K. Wohrer, R. Fosse, M. Chabot, D. Gardes, and C. Champion, Journal of Physics B-Atomic Molecular and Optical Physics **33** (20), 4469 (2000).
[35] R. Fossé, Thesis, Université Pierre et Marie Curie Paris VI, unpublished (2000).
[36] K. Wohrer and R. L. Watson, Physical Review A **48** (6), 4784 (1993).
[37] M. Chabot, R. Fosse, K. Wohrer, D. Gardes, G. Maynard, F. Rabilloud, and F. Spiegelman, European Physical Journal D **14** (1), 5 (2001).
[38] J. M. Hansteen, O. M. Johnsen, and L. Kocbach, Atomic Data and Nuclear Data Tables **15**, 305 (1975).
[39] K. L. Bell, V. Dose, and A. E. Kingston, Journal of Physics B-Atomic and Molecular Physics **2** (831-838) (1969).
[40] G. Herzberg, Electronic spectra and electronic structure of polyatomiques molecules **Van Nostrand Reinhold Company** (1966).
[41] S. Diaz-Tendero, P. A. Hervieux, M. Alcami, and F. Martin, Physical Review A **71** (3) (2005).

[42] S. Diaz-Tendero, G. Sanchez, M. Alcami, F. Martin, P. A. Hervieux, M. Chabot, G. Martinet, P. Desesquelles, F. Mezdari, K. Wohrer-Beroff, S. Della Negra, H. Hamrita, A. Le Padellec, and L. Montagnon, International Journal of Mass Spectrometry **252** (2), 126 (2006).

[43] F. Mezdari, PHD thesis, Université Pierre et Marie Curie Paris VI (unpublished) (2005); M. Chabot, F. Mezdari, G. Martinet, K. Wohrer-Béroff, S. DellaNegra, P. Désesquelles, H. Hamrita, A. LePadellec, and L. Montagnon, Proceedings of the XXIV International conference on Photonic, Electronic and Atomic collisions, 607 (2006).

[44] C. Nicolas, J. N. Shu, D. S. Peterka, M. Hochlaf, L. Poisson, S. R. Leone, and M. Ahmed, Journal of the American Chemical Society **128** (1), 220 (2006).

[45] L. Belau, S. E. Wheeler, B. W. Ticknor, M. Ahmed, S. R. Leone, W. D. Allen, H. F. Schaefer, and M. A. Duncan, Journal of the American Chemical Society **129** (33), 10229 (2007).

[46] R. Ramanathan, J. A. Zimmerman, and J. R. Eyler, Journal of Chemical Physics **98** (10), 7838 (1993).

[47] R. Iftimie, P. Minary, and M. E. Tuckerman, Proceedings of the National Academy of Sciences of the United States of America **102** (19), 6654 (2005).

[48] R. P. A. Bettens and E. Herbst, International Journal of Mass Spectrometry **150**, 321 (1995).

[49] H. Y. Lee, V. V. Kislov, S. H. Lin, A. M. Mebel, and D. M. Neumark, Chemistry-a European Journal **9** (3), 726 (2003).

[50] P. A. Hervieux, B. Zarour, J. Hanssen, M. F. Politis, and F. Martin, Journal of Physics B-Atomic Molecular and Optical Physics **34** (16), 3331 (2001).

[51] F. Calvo and P. Parneix, Computational Materials Science **35** (3), 198 (2006).

[52] H. Choi, R. T. Bise, A. A. Hoops, D. H. Mordaunt, and D. M. Neumark, Journal of Physical Chemistry A **104** (10), 2025 (2000).

[53] O. Heber, K. Seiersen, H. Bluhme, A. Svendsen, L. H. Andersen, and L. Maunoury, Physical Review A **73** (2) (2006).

[54] T. Tuna, M. Chabot, K. Béroff, P. Désesquelles, A. LePadellec, T. Pino, N. T. Van-Oanh, L. Lavergne, F. Mezdari, G. Martinet, M. Barat, and B. Lucas, Proceedings of the "Molecules in Space and Laboratory" conference, editors, J.L.Lemaire and F.Combes, to appear (2007).

[55] G. Herzberg and J. W. C. Johns, The Astrophysical journal **158**, 399 (1969).

[56] R. S. Urdahl, Y. H. Bao, and W. M. Jackson, Chemical Physics Letters **178** (4), 425 (1991).

[57] K. M. Ervin, S. Gronert, S. E. Barlow, M. K. Gilles, A. G. Harrison, V. M. Bierbaum, C. H. Depuy, W. C. Lineberger, and G. B. Ellison, Journal of the American Chemical Society **112** (15), 5750 (1990).

[58] A. M. Mebel and R. I. Kaiser, Chemical Physics Letters **360** (1-2), 139 (2002).

[59] J. Zhou, E. Garand, and D. M. Neumark, Journal of Chemical Physics **127** (15) (2007).

[60] NIST, Webbook of chemistry, National Institute of Standards and Technology **n°69** (2005).

[61] C. J. Reid, J. A. Ballantine, S. R. Andrews, and F. M. Harris, Chemical Physics **190** (1), 113 (1995).

[62] K. Norwood and C. Y. Ng, Journal of Chemical Physics **91** (5), 2898 (1989).

[63] M. S. Deleuze, M. G. Giuffreda, J. P. Francois, and L. S. Cederbaum, Journal of Chemical Physics **112** (12), 5325 (2000).

[64] S. Wilson and S. Green, The Astrophysical Journal **240**, 968 (1980).

[65] M. J. Frisch, G. W. Trucks, H. B. Schlegel, G. E. Scuseria, M. A. Robb, J. R. Cheeseman, J. A. Montgomery, T. Vreven, K. N. Kudin, J. C. Burant, J. M. Millam, S. S. Iyengar, J. Tomasi, V. Barone, B. Mennucci, M. Cossi, G. Scalmani, N. Rega, G. A. Petersson, H. Nakatsuji, M. Hada, M. Ehara, K. Toyota, R. Fukuda, J. Hasegawa, M. Ishida, T. Nakajima, Y. Honda, O. Kitao, H. Nakai, M. Klene, X. Li, J. E. Knox, J. Hratchian, B. Cross, C. Adamo, J. Jaramillo, R. Gomperts, R. E. Stratmann, O. Yazyev, A. J. Austin, R. Cammi, C. Pomelli, J. W. Ochterski, P. Y. Ayala, K. Morokuma, G. A. Voth, P. Salvador, J. J. Dannenberg, V. G. Zakrzewski, S. Dapprich, A. D. Daniels, M. C. Strain, O. Farkas, D. K. Malick, A. D. Rabuck, K. Raghavachari, J. B. Foresman, J. V. Ortiz, Q. Cui, A. G. Baboul, S. Clifford, J. Cioslowski, B. B. Stefanov, G. Liu, A. Liashenko, P. Piskorz, K. Komaromi, R. L. Martin, D. J. Fox, T. Keith, M. A. Al-Laham, C. Y. Peng, A. Nanayakkara, M. Challacombe, P. M. W. Gill, B. Johnson, W. Chen, M. W. Wong, C. Gonzalez, and J. A. Pople, Gaussian Revision C.02, Inc.Wallingford CT (2004).

[66] U. Hechtfischer, C. J. Williams, M. Lange, J. Linkemann, D. Schwalm, R. Wester, A. Wolf, and D. Zajfman, Journal of Chemical Physics **117** (19), 8754 (2002).

[67] C. Petrolongo, P. J. Bruna, and S. D. Peyerimhoff, Journal of Chemical physics **74** (8), 4594 (1981).

[68] J. D. Watts and R. J. Bartlett, Journal of Chemical Physics **96** (8), 6073 (1992).

69 C. E. Moore, National Bureau of Standards Reference Data (1971).

70 M. G. Giuffreda, M. S. Deleuze, and J. P. François, Journal of Physical Chemistry A **103**, 5137 (1999).

71 I. Ben-Itzhak, E. Y. Sidky, I. Gertner, Y. Levy, and B. Rosner, International Journal of Mass Spectrometry **192**, 157 (1999).

## FIGURE CAPTION:

**Figure 1 (Color):**

Two-dimensional representation of signals from neutral fragments in the $C_4H^+$-He collision; colors relate to the number of events, see color scale on the right side.

**Figure 2 :**

Measured cross sections in the $C_nH^+$-He collisions as a function of n. Black circles: total ionisation cross sections; open triangles: electronic excitation (dissociative part); grey hexagons: electron capture. Broken line: IAE calculations for ionization (see text).

**Figure 3:**

Measured probabilities of dissociation, P, of $C_nH$ as a function of the number of emitted fragments $N_f$. Open triangles: CH; full triangles: $C_2H$; open squares: $C_3H$; full circles: $C_4H$. Lines are to guide the eye.

**Figure 4:**

Internal energy distribution of $C_4H$ after electron capture. Grey rectangles are obtained by assuming a step function for f(E); the solid line is obtained by using the analytical form $f(E)=E^{a1}\exp(-a_2(E-a_3)^{a4}$ (see text).

**Figure 5:**

Measured branching ratios of dissociation of $C_nH$ for given values of the number of emitted fragments $N_f$. From left to right: $N_f= 2$ to $N_f= 4$; from bottom to top: $C_2H$ to $C_4H$.

**Figure 6:**

Measured probabilities of dissociation, P, of $C_nH^+$ species as a function of the number of emitted fragments $N_f$. Full triangles: $C_2H^+$; open squares: $C_3H^+$; full circles: $C_4H^+$. Lines are to guide the eye.

**Figure 7:**

Measured branching ratios of $C_nH^+$ for given values of the number of emitted fragments $N_f$. From left to right: $N_f= 2$ to $N_f= 4$; from bottom to top: $CH^+$ to $C_4H^+$